\documentclass[prb,twocolumn,superscriptaddress, floatfix,nobibnotes]{revtex4}
\usepackage{amsmath,amssymb,graphicx,epstopdf,color,bm,soul,upgreek, mathtools,siunitx, fixltx2e,comment,ulem}

\begin{document}

\title{Bloch Oscillations of Hybrid Light-Matter Particles in a Waveguide Array}

\author{J. Beierlein}
\affiliation{Technische Physik, Wilhelm-Conrad-R\"ontgen-Research Center for Complex
Material Systems, and  W\"urzburg-Dresden Cluster of Excellence ct.qmat, Universit\"at W\"urzburg, Am Hubland, D-97074 W\"urzburg,
Germany}

\author{O. A. Egorov}
\affiliation{Institute of Condensed Matter Theory and Optics Friedrich-Schiller-Universit\"at Jena, Max-Wien-Platz 1, D-07743 Jena, Germany}

\author{T. H. Harder}%
\affiliation{Technische Physik, Wilhelm-Conrad-R\"ontgen-Research Center for Complex
Material Systems, and  W\"urzburg-Dresden Cluster of Excellence ct.qmat, Universit\"at W\"urzburg, Am Hubland, D-97074 W\"urzburg,
Germany}

\author{P. Gagel}%
\affiliation{Technische Physik, Wilhelm-Conrad-R\"ontgen-Research Center for Complex
Material Systems, and  W\"urzburg-Dresden Cluster of Excellence ct.qmat, Universit\"at W\"urzburg, Am Hubland, D-97074 W\"urzburg,
Germany}

\author{M. Emmerling}%
\affiliation{Technische Physik, Wilhelm-Conrad-R\"ontgen-Research Center for Complex
Material Systems, and  W\"urzburg-Dresden Cluster of Excellence ct.qmat, Universit\"at W\"urzburg, Am Hubland, D-97074 W\"urzburg,
Germany}

\author{C. Schneider}%
\affiliation{Institute of Physics, University of Oldenburg, D-26129 Oldenburg, Germany}

\author{S. H\"ofling}
\affiliation{Technische Physik, Wilhelm-Conrad-R\"ontgen-Research Center for Complex
Material Systems, and  W\"urzburg-Dresden Cluster of Excellence ct.qmat, Universit\"at W\"urzburg, Am Hubland, D-97074 W\"urzburg,
Germany}

\affiliation{SUPA, School of Physics and Astronomy, University of St Andrews, St Andrews
KY16 9SS, United Kingdom}

\author{U. Peschel}
\email{ulf.peschel@uni-jena.de}
\affiliation{Institute of Condensed Matter Theory and Optics Friedrich-Schiller-Universit\"at Jena, Max-Wien-Platz 1, D-07743 Jena, Germany}

\author{S. Klembt}
\email{sebastian.klembt@ physik.uni-wuerzburg.de.}
\affiliation{Technische Physik, Wilhelm-Conrad-R\"ontgen-Research Center for Complex
Material Systems, and  W\"urzburg-Dresden Cluster of Excellence ct.qmat, Universit\"at W\"urzburg, Am Hubland, D-97074 W\"urzburg,
Germany}

\begin{abstract}
Bloch oscillations are a phenomenon well known from quantum mechanics where electrons in a lattice experience an oscillatory motion in the presence of an electric field gradient. Here, we report on Bloch oscillations of hybrid light-matter particles, called exciton-polaritons, being confined in an array of coupled microcavity waveguides. To this end, we carefully design the waveguides, widths and their mutual couplings such that a constant energy gradient is induced perpendicular to the direction of motion of the propagating exciton-polaritons. This technique allows us to directly observe and study Bloch oscillations in real- and momentum-space. Furthermore, we support our experimental findings by numerical simulations based on a modified Gross-Pitaevskii approach. Our work provides an important transfer of basic concepts of quantum mechanics to integrated solid state devices, using quantum fluids of light.
\end{abstract}

\maketitle
\section{INTRODUCTION}

Already in the early days of quantum mechanics, the motion of particles in a periodic potential under the action of a constant force was investigated extensively to understand electrical currents induced by external fields acting on solid materials. But in contradiction to the classical expectation, particles did not follow the direction of the driving force but instead performed an oscillatory motion, so-called Bloch oscillations \cite{Bloch1929}. It was soon understood that those Bloch oscillations occur because the external force causes the particle to gain momentum, thus changing their location inside the Brillouin zone. This motion in momentum space continues going even beyond the edge of the Brillouin zone. Since the band structure is periodic, the initial field distribution is recovered after one crossing of the Brillouin zone. Therefore, one observes an oscillatory motion but no net shift of the particle in real space. 
In order to explain the motion of electrons in a crystalline lattice under the action of a dc electric field, this single band picture had to be extended to take into account the coupling to other bands, an effect known as Zener tunneling \cite{Zener1934}. While originally predicted in the context of electrons in crystals, Bloch oscillations were also extensively investigated in different physical systems, including electrons in semiconductor superlattices \cite{Feldmann1992}, cold atoms in optical lattices \cite{Dahan1996,Morsch2001,Geiger2018} and phonons in acoustic microcavities \cite{Lanzillotti-Kimura2010}.\\
Additionally, many effects originally predicted in solid-state physics have been observed in optics by monitoring light propagation in photonic lattices.
Also one-dimensional optical Bloch oscillations were observed in arrays of coupled dielectric waveguides with a transversely superimposed linear ramp of the effective index \cite{Pertsch1999, Morandotti1999a}. Later, similar experiments were also performed in silica \cite{Szameit2010} and plasmonic \cite{Block2014} waveguide arrays. In all cases, a periodic distribution of the refractive index plays the role of the crystalline potential, and the  gradient of the effective indices of the waveguides acts similar to an external force in a quantum system. It causes the beam to move across the waveguide array where it experiences Bragg reflection on the high-index and total internal reflection on the low-index side of the structure, resulting in an optical analogue of Bloch oscillations.
When many-body effects induce nonlinear dynamics, physics becomes even richer. Several experiments with quantum fluids of light, using periodic potentials, have observed such intriguing phenomena as the Mott insulator transition \cite{Greiner2002} or a breakdown of superfluidity in atomic Bose-Einstein condensates \cite{Burger2001}.
Microcavity exciton-polaritons (polaritons) as hybrid light-matter particles, arise from the strong coupling of a photonic cavity mode (light) and a quantum well exciton (matter). They inherent a unique combination of properties from their part-light, part-matter nature, most notably a small effective mass inherited from the photonic component, as well as the ability to interact from the excitonic component. Consequently, they have been described as \textit{quantum fluids of light}\cite{Carusotto2013}. After the first demonstration of the strong-coupling regime \cite{Weisbuch1992}, and the possibility to witness a phase transition to a macroscopic occupation of a single polariton ground state, referred to as polariton condensation \cite{Kasprzak2006}, researchers have soon begun to study propagating polaritons in different waveguiding settings. 

\section{FREE PROPAGATION OF MICROCAVITY POLARITONS AND PROPAGATION WITHIN WAVEGUIDES}

Propagating exciton-polaritons show one of the most spectacular phenomena of quantum fluids, namely superfluidity \cite{Amo2009a,Amo2009b}, manifesting itself as the suppression of scattering from defects when the flow velocity is less than the speed of sound in the fluid. For larger flow velocities the perturbation induced by the defect gives rise to turbulent emission of quantized vortices and to the nucleation of solitons \cite{Amo2011,Flayac2011a,Wouters2010,Lagoudakis2008}. Further experimental investigations of the coherently driven semiconductor microcavities provide the evidence of polariton droplets moving on top of the condensate background with a velocity as high as \SI{1.2d6}{m/s} \,(cf. refs.\cite{Flayac2011a,Sich2012}). Furthermore, it was found that the strong repulsive force between microcavity polaritons, originating from exciton-exciton interaction results in a substantial nonlinearity, which causes a density dependent blue shift of the exciton resonance \cite{Ciuti2003} and can be exploited to force  polaritons to propagate along a single waveguide \cite{Wertz2012}, or waveguide coupler devices \cite{Beierlein2020}. Quite recently, the realisation of polariton waveguides using perovskite materials at room temperature has rekindled the interest in such devices \cite{Su2018}.\\
So far, very few papers have been devoted to theoretical studies of Bloch oscillations of polaritons in solid-state devises. However, the required conditions and stability of Bloch oscillations have been theoretically estimated for a one-dimensional lattice embedded into a semiconductor resonator operating in the strong light-matter coupling regime \cite{Flayac2011b,Flayac2011c}.
Here, we implement a one-dimensional waveguide array for exciton-polaritons, supporting a significant gradient perpendicular to the direction of motion, unequivocally demonstrating polariton Bloch oscillations. Additionally, we develop a theoretical framework, to model the experiments presented in this work.
\indent 
Creating a desired coupling between waveguides has been a challenge in various material platforms. In this work, we take advantage of the etch-and-overgrowth technique which allows for a large range and precise control of the coupling strength between two adjacent trapping potentials.

\begin{figure}[t!]
\centering
\includegraphics[width=1\linewidth]{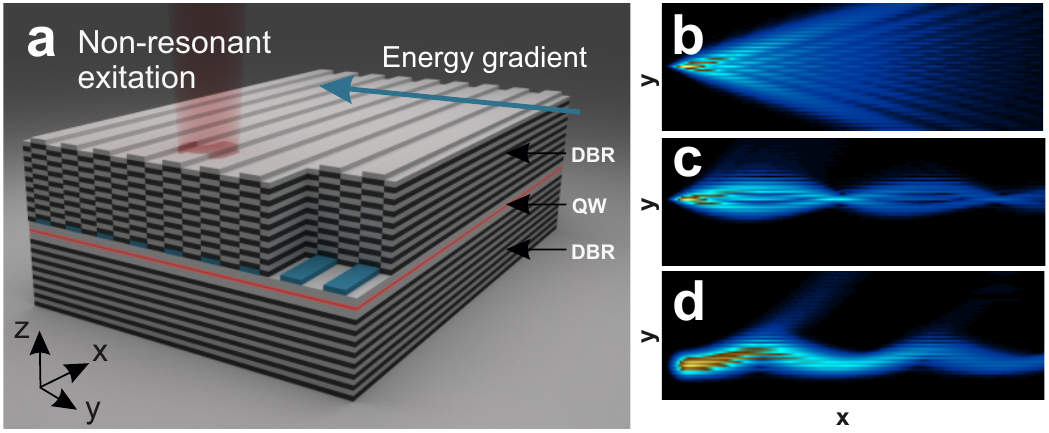} 
\caption{(a) Schematic of the etch-and-overgrowth microcavity waveguide array investigated here. The red layer indicates the quantum well stacks placed between two DBR mirrors. The blue areas indicate the waveguide structures etched into the cavity layer. (b) Simulated propagation of a polariton condensate, along the $x$-direction, excited in one waveguide in a homogeneous waveguide array with no energy gradient. The characteristic fan-like pattern of discrete diffraction is observed. (c,d) Simulated propagation of a condensate inside a coupled waveguide array with a potential gradient in $y$-direction and excited in a single waveguide or multiple waveguides, respectively. While for the excitation in a single waveguide the intensity undergoes a refocusing of the intensity, for multiple waveguide excitation the intensity shows a sinusoidal oscillation pattern.
}
\label{fig1}
\end{figure}

\section{EXPERIMENTAL SETUP}
The specific sample designed and investigated for this work consists of 37 bottom and 32 top Al\textsubscript{0.2}Ga\textsubscript{0.8}As/AlAs mirror pairs. As active material two stacks of four GaAs quantum wells with an exciton energy of $E\textsubscript{X} = \SI{1.614}{\electronvolt}$ and a width of \SI{7}{\nano\meter}, were embedded in the structure. The stacks were positioned in the antinode as well as the first bootom mirror pair of the AlAs $\lambda$/2-cavity to maximize their overlap with the electric field.
In this approach, the molecular beam epitaxial growth is stopped after the bottom distributed Bragg-reflector (DBR) and the cavity layers have been  grown. Subsequently, waveguide arrays are patterned into a spacer layer using an electron-beam lithography process followed by a wet etch step. During this etch step the actual cavity length is reduced resulting in a blue shift of the etched area. This acts as a potential barrier between the patterned waveguides, where the barrier height can be tuned accurately with an etch depth of nanometer precision. To complete the vertical confinement another DBR is epitaxially grown on top. In the samples employed in this study the etch depth is of the order of $\sim$\,\SI{10}{\nano\meter} resulting in a confinement potential of $\sim$\,\SI{6.7}{\milli\electronvolt}. A rendered schematic of the sample is shown in Fig.~\ref{fig1}(a), here the patterning of the cavity is highlighted without the top mirror. It has been previously shown that the method is well suited to create high-quality polariton waveguides \cite{Winkler2017}, as well as its suitability to accurately control coupling between adjacent sites \cite{Harder2020}.
The Rabi splitting of the sample was determined using a white light reflectivity measurement at \SI{10}{\kelvin} to \SI{11.5}{\milli\electronvolt}. By measuring the photoluminescence (PL) of a photonic detuned area of the sample the quality factor of the wafer could be extracted to Q $\approx 7500 $.\\
The spectroscopic results presented in this work were measured using momentum-resolved PL spectroscopy. The sample was placed inside a liquid helium flow cryostat and kept at $T$ = \SI{4.2}{\kelvin}. Laser excitation was provided using a tuneable continuous wave Ti:sapphire laser, set to the energy of \SI{1.684}{\electronvolt} corresponding to the first high-energy Bragg minimum. The pump spot was focused via a microscope objective with a numerical aperture $NA = 0.42$ to a diameter of $\sim \,$\SI{3}{\micro\meter}. The detection path of the setup allowed for real-space as well as momentum-resolved measurements. By motorizing a lens in the detection path it is possible to make a full tomography of the real space and emission energy $E_{PL}(x,y)$.  By using a real space focus plane, the PL could be spatially selected before conversion into Fourier space. The emission is then energy-resolved by a Czerny-Turner monochromator and detected on a 1024$\times$1024 pixel charge-coupled device camera, cooled down to \SI{-75}{\celsius} by a Peltier cooler.\\

\section{ENGINEERING OF COUPLED POLARITON WAVEGUIDE ARRAYS}

\begin{figure}[!]
\centering
\includegraphics[width=\linewidth]{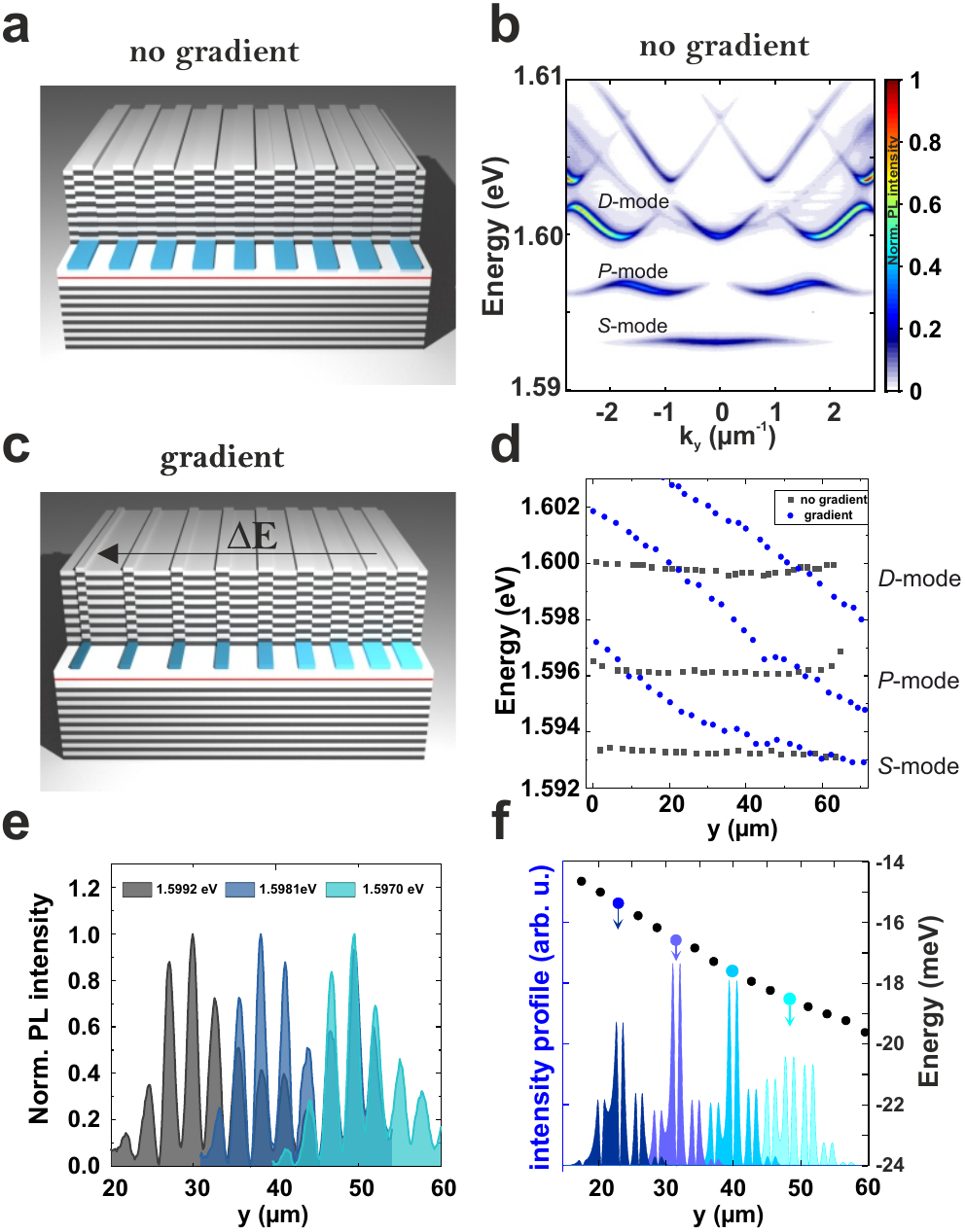} 
\caption{(a,c) Rendered schematics of waveguide arrays under investigaton. While (a) has  constant waveguide widths, (c) has an engineered energy gradient in the $y$-direction, by adjusting the waveguide widths. In (b) the dispersion along the $k_{y}$ direction of the homogeneous waveguide array (a) is depicted by a momentum-resolved energy spectrum. Measured  spatially-resolved energy spectra of both arrays are presented in (d). Each waveguide mode gives rise to the formation of a flat band in case of the homogeneous (a) and of a ladder like structure - the so-called Wannier-Stark ladder - in case of the inhomogeneous (c) array. (e,f) Experimentally (e) and numerically (f) determined $y$-dependent spatial distributions of Wannier-Stark states with different energies along the $y$-direction. (f) Theoretical calculations of different Wannier states inside the gradient-engineered waveguide array (d). }
\label{fig2}
\end{figure}

In Fig. \ref{fig1}(a), a sketch of a homogeneous waveguide array (blue) patterned into the cavity layer is depicted. The propagating polariton condensate is formed using non-resonant laser excitation. The gradient needed to invoke Bloch oscillations in this waveguide array is oriented in $y$-direction, perpendicular to the propagation direction. While the use of Gross-Pitaevskii models has been fairly well established for exciton-polaritons in the past, we have successfully expanded these models to take into account sophisticated lattice potentials \cite{Harder2020b,Harder2020c}.
By tuning the gradient, coupling properties and excitation conditions, different regimes of propagation patterns can be achieved. Such patterns for a propagating polariton condensate are calculated in Figs.~\ref{fig1}(b-d). In Fig.~\ref{fig1}(b), the propagation for a homogeneous (vanishing gradient) waveguide array, where a fan-like pattern typical for discrete scattering emerges, is displayed. Note that due to the forward propagation of the condensate a fast temporal evolution is transformed into a spatial distribution which can be detected rather easily. Hence, we expect to see Bloch oscillations to unfold in real space, which is also confirmed by respective simulations. In Figs.~\ref{fig1}(c,d) an energy gradient in the $y$-direction is applied with excitation on a single waveguide and multiple waveguides, respectively. For a single waveguide excitation, the intensity couples to the neighbouring waveguides before being refocused to a single spot. This behaviour is a typical fingerprint of Bloch oscillations and of the transformation of extended Bloch waves into localized Wannier states due to the action of the linearly growing potential. As all energy eigenstates are equally spaced in a so-called Wannier-Stark ladder, every excitation must recover after a finite evolution time or propagation length, respectively. Hence, if a single waveguide is excited (see Fig.~\ref{fig1}(c)) the field must refocus after a Bloch period. In contrast, pronounced sinusoidal oscillations are observed for an excitation of several waveguides (see Fig.~\ref{fig1}(d)).\\
To achieve a sizeable and controllable gradient in the energy landscape in a polariton system, either the excitonic or photonic component of the exciton-polariton can be altered. 
Here, the advantage of the etch-and-overgrowth technique comes into play, allowing control of the width and position of the waveguides with remarkable e-beam lithography precision while preserving an extended mode distribution due to the highly controllable and shallow photonic barrier in comparison to traditional dry etch techniques, e.g. employed in ref. \cite{Wertz2010}.

Fig.~\ref{fig2} displays different waveguide arrays under investigation including experimentally detected momentum-resolved energy spectra. We start with a homogeneous array (without gradient) displayed in Fig.~\ref{fig2}(a) with a lattice constant of $a=$ \SI{2.8}{\micro\meter} and a waveguide width of $d=$ \SI{2}{\micro\meter}. Its dispersion along the $y$-direction is depicted in Fig.~\ref{fig2}(b). Here, the coupling between the photonic modes results in the formation of a band structure where each waveguide mode gives rise to an individual band. The curvature of each band is determined by the coupling between the waveguides. With increasing mode number and energy, the extension of the modes into the barrier increases, therefore enhancing the coupling. While the two lowermost $S-$ and $P-$modes are nicely confined, the D-mode is already at the same energy level as the barrier, indicated by the low-intensity parabola at $E_{\text{barr}} (k_y=0) \approx  1.601$\,eV. Following the linear wave approximation, propagation constants of  waveguide modes are unambiguously related to their energies. In order to create the required transverse gradient, we therefore modified the widths of the waveguides such that the propagation constants of the ground modes for the fixed energy decreased accordingly.
By keeping the center-to-center lattice constant at $a= \SI{2.8}{\micro\meter}$ but gradually increasing the width of the waveguides the energy levels were tuned similar to a ladder (see Fig.~\ref{fig2}(c)). The resulting spatially-resolved energy landscape for the different modes and for both arrays is displayed in Fig.~\ref{fig2}(d) showing a linear gradient, in comparison to the very small gradient of the homogeneous lattice originating from the epitaxial growth \cite{Comment_Grad}. 
Respective eigenstates are represented in Fig.~\ref{fig2}(e) by spectrally and spatially resolving the emitted PL of the \textit{P}-mode  (see Fig.~\ref{fig2}(d)). Numerically determined Wannier-Stark states for such a gradient-engineered waveguide array are depicted in Fig.~\ref{fig2}(f). We find all field profiles to  be completely localized and to have approximately the same shape. States of different energies differ only with respect to their position on the energetically inclined lattice.
If several of these states are excited, their spatial extension  defines the elongation of resulting Bloch oscillations.
It is worth pointing out, that the experimental images are limited by the finite spatial resolution of the setup, which doesn't allow us to resolve the \textit{P}-mode profile (see Fig.~\ref{fig2}(f)).
Besides, our array is slightly spatially inhomogenous due to the varying width of the waveguides resulting in an artificial compression or stretching of respective field profiles. A certain spread of wave vectors in propagation direction $x$ causes an additional blurring. Keeping these experimental limitations in mind, there is nonetheless a very good agreement between the experimentally determined Wannier states and the theoretical model.
The induced gradient of the \textit{P}-mode can be fitted to roughly \SI{260}{\micro\electronvolt} per guide or $\sim \SI{100}{\micro\electronvolt \per\micro\metre}$, approximately two orders of magnitude larger than the gradient naturally occuring in the epitaxial microcavity growth \cite{Comment_Grad}.

\section{Optical Study of Polariton Bloch Oscillations}

\begin{figure*}[t!]
\centering
\includegraphics[width=\linewidth]{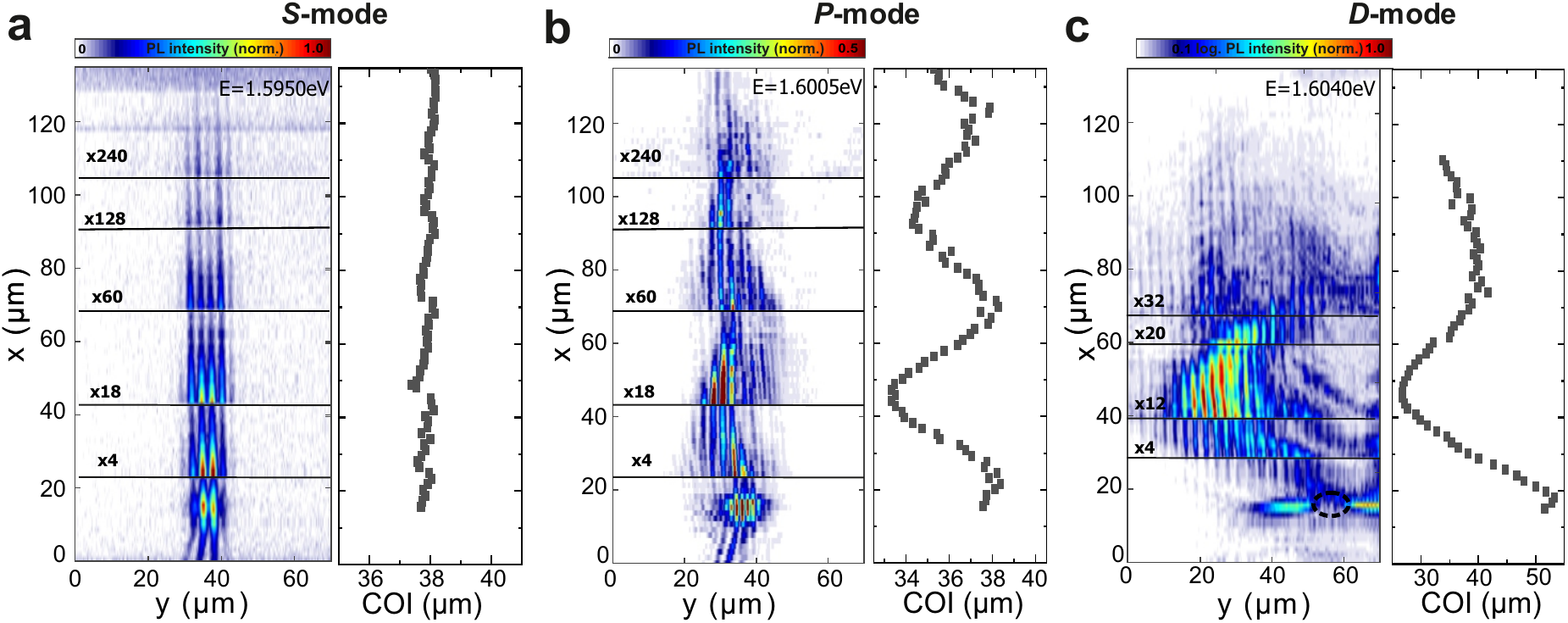} 
\caption{(a) Energy-resolved emission of a \textit{S}-mode condensate propagating along the waveguide array with a gradient but without sufficient coupling to other waveguides, with the corresponding COI.(b) Propagating \textit{P}-mode polariton condensate showing a clear oscillation in the $y$-direction. Its COI highlights the oscillation with a period of  $\sim$\SI{50}{\micro\metre}. (c) Energy-resolved emission of a \textit{D}-mode propagating in an arch like pattern, underlined by the COI.}
\label{fig3}
\end{figure*}

In order to study polariton flow in the waveguide array, the sample was excited with a focused laser spot. When the power is increased, a distinct condensation threshold is reached at $P_{th}= \SI{3}{\kilo\watt\per\square\centi\metre}$. Due to the interaction-induced blue shift caused by their excitonic fraction, polaritons are expelled from the excitation spot due to Coulomb interactions. This can be verified by a localized blue shift at the excitation spot accompanied by outflowing polaritons with a finite momentum along the waveguide direction \cite{Wertz2010,Rozas2020}. In Figs.~\ref{fig3}(a-c), propagating condensates originating from the \textit{S}-, \textit{P}- and  \textit{D}-modes are depicted. The images were obtained by plotting the PL at respective energies $E_{S}=\SI{1.5950}{\electronvolt}$,  $E_{P}=\SI{1.6005}{\electronvolt}$ and $E_{D}=\SI{1.6040}{\electronvolt}$. Note that, due to the blue shift induced by the condensate, the respective energies are upshifted by about 2.5meV compared with the data displayed in Fig.~\ref{fig2}(d).  For the \textit{S} and \textit{P}-modes, the array was excited in the center, while for Fig.~\ref{fig3}(c) the laser was shifted to the indicated position (see dashed black cirlce). Due to the dissipative nature of the polaritons, the PL signal in  Fig.~\ref{fig3}(a-c) had to be attenuated using a neutral-density filter to show the propagation along the entire structure. To analyze the intensity distribution we define the Center of Intensity (COI) described by :

\begin{equation}
COI(x) = \frac{\sum_{y} I_{x,y}*y }{\sum_{y} I_{x,y}} 
\label{eq1}
\end{equation}
along the propagation direction. Here, $x$ and $y$ corresponds to the spatial position, and $I_{x,y}$ to the intensity of a given pixel. In Fig.~\ref{fig3}(a), the condensate spreads only sparsly to the neighbouring waveguides in $y$-direction and shows almost no oscillations. This behaviour is no surprise as the $S$-mode is subject to the strongest confinement, resulting in a vanishing coupling to adjacent waveguides. Therefore the width of the propagating condensate of a few waveguides is dominated by the size of the excitation spot. Consequentially, for the \textit{S}-mode the COI results in almost a straight line. Opposed to the $S$-mode, the $P$-mode depicted in Fig.~\ref{fig3}(b) unequivocally shows an oscillation in the $y$-direction, perpendicular to the direction of motion. The COI highlights this behavior and allows to determine an oscillation amplitude $A=$\,\SI{2.7}{\micro\meter}. Hence, about three waveguides are involved in the oscillation which is in agreement with the extension of Wannier-Stark states displayed in Fig.~\ref{fig2}(e) and (f). The observed period of oscillations of $l_{\lambda}\sim\,$\SI{50}{\micro\meter} together with the energy spacing per guide, displayed in Fig.~\ref{fig2}(d), hints to a speed of the propagating \textit{P}-mode condensate of roughly $\sim \SI{3.1d6}{\metre\per\second}$.

While the engineered gradient is similar for all photonic bands (see Fig.~\ref{fig2}(d)), the coupling increases with the photonic band energy, due to lower confinement of respective Bloch modes.  Consequently the elongation of \textit{P}-mode Bloch oscillations increases. 
In the experiment we simultaneously excited two to three waveguides and consequently the resulting condensate covers only the central part of the Brillouin zone. This is the situation displayed in Fig.~\ref{fig1}(d) and different from the point-like excitation displayed in Fig.~\ref{fig1}(c). The latter one corresponds to a simultaneous excitation of all Bloch waves of the array including those having a negative effective mass and thus initially moving opposite to the direction of the gradient. 
If the coupling is increased even further by switching to \textit{D}-modes, an oscillation pattern with much larger amplitude of $A=$\,\SI{11}{\micro\meter} and oscillation period $l_{\lambda}\sim$\,\SI{65}{\micro\meter} is observed which hints to a sligthly higher speed of the \textit{D}-mode condensate of  \SI{4.1d6}{\metre\per\second}\\


These results demonstrate that for sufficient coupling between the waveguides while a sizeable gradient perpendicular to the direction of motion is present, the polaritons undergo distinct Bloch oscillations.

\begin{figure}[t!]
\centering
\includegraphics[width=0.98\linewidth]{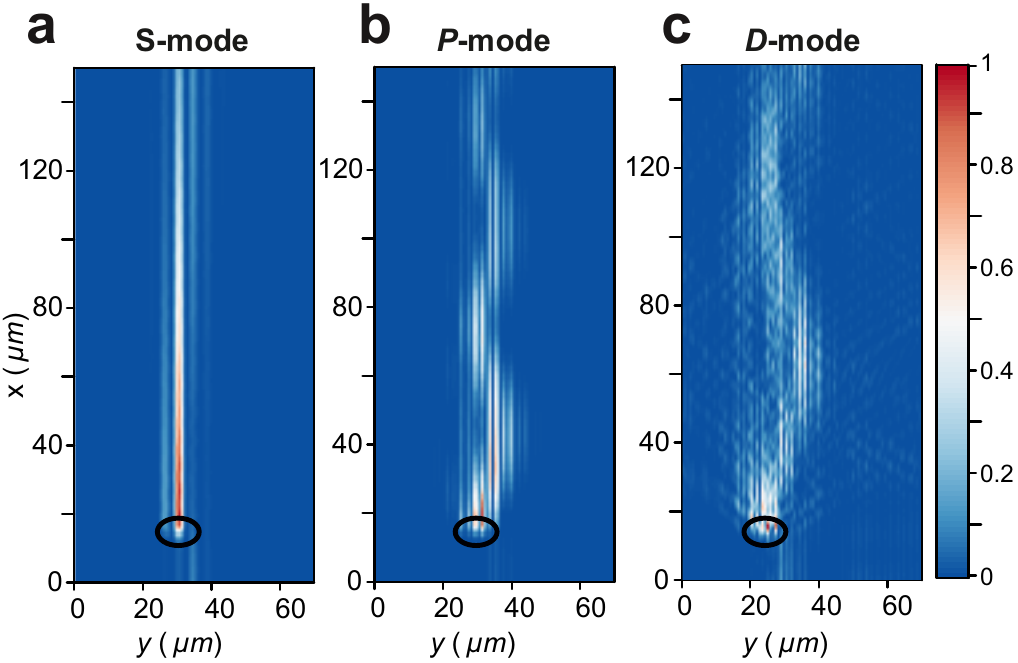}
\caption{Simulated propagation of a condensate excited in an array with a potential landscape with an increasing width of waveguides in $y$-direction. (a) Excitation of \textit{S}-modes of three neighbouring waveguides by a coherent pump of momenta $k_p=$\SI{2}{\per\micro\meter} and frequency equivalent energy $\hbar \Delta_\text{e}=$ \SI{-15}{\milli\electronvolt}. A negligible coupling to the neighbouring waveguides can be observed, leaving the main part of signal on the center waveguide. (b) By exciting multiple waveguides in the \textit{P}-mode the condensate undergoes periodic behaviour along the $y$-axis, since the coupling within the \textit{P}-band is sufficient.
(d) Multiple excitation of the \textit{D}-modes of the array by a coherent pump of momenta $k_p=$ \SI{2}{\per\micro\meter} and frequency equivalent energy $\hbar \Delta_\text{e}=$\SI{-9}{\milli\electronvolt}.   }                 
\label{fig4}
\end{figure}

\section{Numerical study of Bloch Oscillations using a modified Gross-Pitaevskii model}

To underpin these experimental results we performed numerical calculations of exciton-polariton dynamics in semiconductor microcavities. Neglecting polarization effects one obtains two coupled Schr\"odinger equations for the intracavity photonic field $\Psi_\text{c}$ and coherent excitons  $\Psi_\text{e}$ \cite{Amo2009a, Carusotto2013} given as
\begin{equation}\begin{split}
\label{eq:phot}
\partial _t & {\Psi_\text{c} }    - \frac{i\hbar}{2m_{\text{c}}}  \nabla_{x,y} ^2{\Psi_\text{c} } + i V(y) {\Psi_\text{c} } \\
& + \left[ \gamma_\text{c}  - i \Delta_\text{c} \right]{\Psi_\text{c} }    = i{\Omega _\text{R}{\Psi_\text{e} } + E (x,y) e^{ik_\text{p}x}},
\end{split}
\end{equation}
\begin{equation}\begin{split}
\label{eq:exc}
\partial _t{\Psi_\text{e} }  - \frac{i\hbar}{2m_{\text{ex}}}  \nabla _{x,y} ^2{\Psi_\text{e} } + \left[ {\gamma_\text{e}  - i{ \Delta_\text{e}  }} \right]{\Psi_\text{e}} = i{\Omega _\text{R}}{\Psi_\text{c}  \,  .}
\end{split}
\end{equation}
The complex amplitudes $\Psi_\text{c}$ and $\Psi_\text{e}$ are obtained by averaging related creation or annihilation operators.  The effective photon mass in the planar region is given by $m_\text{c}=42.33 \times 10^{-6} \, m_\text{e}$ where $m_\text{e}$ is the free electron mass. The effective mass of excitons $m_\text{ex}\approx 10^5 \, m_\text{c}$ is so high that independent exciton propagation can be neglected.   The coupling strength between intracavity photons and excitons  $\Omega_{\text{R}}$ defines the Rabi splitting $2\hbar \Omega_{\text{R}}=11.5\,\text{meV}$. The parameters $\Delta_\text{c,e} = {\omega _\text{p}-\omega _\text{c,e}} $ account for the frequency detunings of the operating frequency $\omega _\text{p}$ from the cavity $\omega _\text{c}$ and excitonic $\omega _\text{e}$ resonances, respectively. Then the exciton photon detuning of the device is given by $ \hbar \omega_{e} - \hbar \omega_{c}=\hbar \Delta_{c} - \hbar \Delta_{e}= - 15.9 \, \text{meV} $.  $\gamma_\text{c}$ = $\gamma_\text{e}$= \SI{0.01}{\milli\electronvolt} are the cavity photon and exciton damping constants. 
An external photonic potential $V(x,y)$ defines the waveguide geometry induced by structuring of the planar microcavity. In our modelling, a separate waveguide profile of the array is given by a super-Gauss $V(y)=V_0\,(1-\text{exp}(-y^{24}/s^{24}))$ with the potential depth $\hbar V_0= 6\, \text{meV}$ and waveguide width $s$.

Within the modelling we skip details of condensation dynamics and focus on the propagation of polaritons in the waveguide array. Therefore, the initial condensate with a well defined frequency and momentum is artificially launched by a localized coherent pump term $E_\text{p} (x,y)$ with a frequency $\omega_p$ and with a momentum $k_\text{p}$. 
Fig.~\ref{fig4} shows the results of numerical simulations of polariton propagation dynamics in the potential landscape closely related to the experimental configuration. The appropriate choice of the momentum of the launched beam ($k_p$)  allows for a direct excitation of the desired waveguide mode. Similar to the experimental results discussed above, exciton-polaritons launched into the \textit{S}-mode do not couple to neighbouring waveguides, thus keeping their energy in the excitation guides (see Fig.~\ref{fig4}(a)). In contrast, typical Bloch oscillation dynamics occur after excitation of a higher waveguide mode (\textit{P}-mode), as it is shown in Fig.~\ref{fig4}(b). 
Even more pronounced periodic motion of the wave packets can be recognised for an excitation of (\textit{D}-modes) as demonstrated in Fig.~\ref{fig4}(c) in full agreement with the experiments displayed in Fig.~\ref{fig3}.

\section{Momentum Distributions of Bloch Oscillations: Experiment and Theory}

\begin{figure} 
\centering
\includegraphics[width=\linewidth]{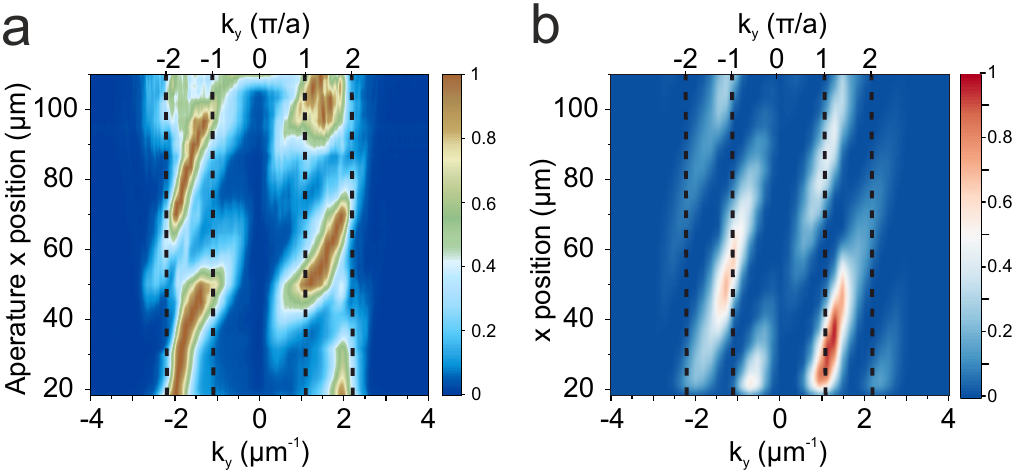} 
\caption{(a,b) $k_y$ dispersion of the propagating condensate at various spatial positions. The periodic change of the wavevector as well as the characteristic Bragg reflection at the edge of the Brillouin zone (black dashed lines) is shown for the \textit{P}-mode oscillation of Fig.~\ref{fig3}(b), at $E_{P}=\SI{1.6005}{\electronvolt}$, experimentally (a) and the  theoretical calculation of Fig.~\ref{fig4}(b) in (b), respectively. The Brillouin zone is given by $\pi/a$, with $a=\SI{2.8}{\micro\metre}$.  }
\label{fig5}
\end{figure}

The characteristic sweeping over the whole Brillouin zone during the Bloch oscillations can be illustrated by performing a spatial Fourier transformation of the propagating polariton field emitted from the cavity. In Fig.~\ref{fig5}(a) we show the experimental momentum dispersion of the condensate from in $y$-direction for different positions along the propagation direction $x$, which corresponds to the far field of the field distribution displayed Fig.~\ref{fig3}(b). In order to directly image the dispersion, a slice of the real space image was cut out by an optical aperture,that was mounted on a motorized stage to tune the $x$-position. Next, the resulting slice of the optical field was imaged to infinity  corresponding to a spatial Fourier transformation. Finally the energy of the mode of interest was selected using the monochromator. 
While experimental and numerical data agree very well (see Fig.~\ref{fig5}(a) and (b)), one has to mention that the displayed images are not a one-to-one illustration of the occupation of the Brillouin zone. First, we image the zeroth and the  +/- first diffracted order corresponding to a three time repetition of the periodic Brillouin zone the boundaries of which are marked by black dashed lines. Second, the whole image is overlaid with the Fourier transform of the \textit{P}-mode profile, covering the whole displayed Fourier space. Note, that it has a zero in its center thus rendering the lowest order Brillouin zone almost invisible as well as being limited by the dispersion of the cavity for higher wave vectors.
Nonetheless, the expected dynamics of the momentum are clearly visible.

During propagation the wavevector of the \textit{P}-mode shifts towards positive $k_{y}$. Hence, due to the action of the gradient the field gains momentum before undergoing a Bragg reflection at the edge of the Brillouin zone. The latter process causes the beam to deflect back thus performing a Bloch oscillation. This characteristic behavior has been directly imaged in cold atoms in one-dimensional optical lattices before \cite{Atala2013} and now unambiguously supports the demonstration of Bloch oscillation in a polaritonic quantum fluid of light. Consistently, the numerical simulation shown in Fig.~\ref{fig5}(b) of our system agree and underline our results.

\section{Conclusion}

In conclusion, we have introduced a general approach to study Bloch oscillations of exciton-polaritons in semiconductor waveguide arrays. Doing so we demonstrate the first Bloch oscillations of a propagating exciton-polariton condensate in a waveguide array. By using an etch-and-overgrowth approach we are able not only to create a sufficient coupling between the physically separate waveguides, but also to induce a potential gradient for the quantum fluid. This approach allows for highly tuneable system parameters such as waveguide width and coupling strength laying the foundation for future experiments requiring precise system control, for example Zener tunneling or effects in Parity-Time ($\mathcal{PT}$) symmetric potentials, involving gain and loss.
In addition to that, this work opens the way towards the implementation of other physical effects like self-localization \cite{Anderson1958,Pertsch2004,Schwartz2007,Sheinfux2017}, implementation of Floquet theory \cite{Lindner2011,Rechtsman2013,Clark2019} as well as topological protection \cite{Hafezi2013,Khanikaev2013,Blanco-Redondo2016,Barik2018,Klembt2018,Upreti2020} in a driven-dissipative waveguide system featuring strong interactions as well as a large nonlinearity.\\


\begin{acknowledgments}
The W\"urzburg and Jena group acknowledge financial support within the DFG projects PE 523/18-1 and KL3124/2-1. The W\"urzburg group acknowledges financial support by the German Research Foundation (DFG) under Germany’s Excellence Strategy–EXC2147 “ct.qmat” (project id 390858490).
S.H. also acknowledges support by the EPSRC “Hybrid Polaritonics” grant (EP/M025330/1). T.H.H. and S.H. acknowledge funding by the doctoral training program Elitenetzwerk Bayern Graduate School “Topological insulators” (Tols 836315).  T.H.H. acknowledges support by the German Academic Scholarship Foundation.

\end{acknowledgments}


\begin{thebibliography}{99}
\bibitem{Comment_Grad} Molecular beam epitaxy typically results in a radial thickness gradient from the center to the edge of a waver. This is the reason why the cavity exhibits a (small) blue shift with increasing radial distance from the wafer center. Typical gradients are of the order of \SI{2}{\micro\electronvolt\per\micro\metre} and are thus alone insufficient for the observation of Bloch oscillations.
\end{thebibliography}
\end{document}